\documentclass[prl,letterpaper,twocolumn,showpacs,showkeys,lengthcheck,
               floatfix,nofootinbib,preprintnumbers,superscriptaddress]{revtex4-1} 

\usepackage{amsmath,amssymb,amsfonts,graphicx}
\usepackage{bm}
\usepackage{hyperref}
\usepackage{slashed}
\usepackage[caption=false]{subfig}
\usepackage[svgnames]{xcolor}
\usepackage[english]{babel}
\usepackage{blindtext}
\usepackage{microtype}
\usepackage{tikz}

\RequirePackage{slashed}

\def\be{\begin{equation}}
\def\ee{\end{equation}}

\begin{document}


\title{Octet Spin Fractions and the Proton Spin Problem}

\author{P.~E.~Shanahan}
\affiliation{CSSM and ARC Centre of Excellence for Particle Physics at the Tera-scale,\\ 
School of Chemistry and Physics, 
University of Adelaide, Adelaide SA 5005, Australia
}
\author{A.~W.~Thomas}
\affiliation{CSSM and ARC Centre of Excellence for Particle Physics at the Tera-scale,\\
School of Chemistry and Physics,
University of Adelaide, Adelaide SA 5005, Australia
}
\author{K.~Tsushima}
\affiliation{CSSM and ARC Centre of Excellence for Particle Physics at the Tera-scale,\\
School of Chemistry and Physics,
University of Adelaide, Adelaide SA 5005, Australia
}
\author{R.~D.~Young}
\affiliation{CSSM and ARC Centre of Excellence for Particle Physics at the Tera-scale,\\
School of Chemistry and Physics,
University of Adelaide, Adelaide SA 5005, Australia
}
\author{F.~Myhrer}
\affiliation{Department of Physics, University of South Carolina,
             Columbia SC, USA}
\begin{abstract}
The relatively small fraction of the spin of the proton carried by its 
quarks presents a major challenge to our understanding of the strong 
interaction. Traditional efforts to explore this problem have involved 
new and imaginative experiments and QCD based studies of the nucleon. 
We propose a new approach to the problem which exploits recent advances 
in lattice QCD. In particular, we extract values for the spin carried 
by the quarks in other members of the baryon octet in order to see 
whether the suppression observed for the proton is a general property 
or depends significantly on the baryon structure. 
We compare these results 
with the values for the spin fractions calculated within a model that 
includes the effects of confinement, 
relativity, gluon exchange currents and the meson 
cloud required by chiral symmetry, finding a very satisfactory level 
of agreement given the 
precision currently attainable.
\end{abstract}


\maketitle


There have been decades of careful experimental investigation since 
the original discovery of the so-called proton spin crisis by the 
European Muon Collaboration 
(EMC)~\cite{Ashman:1987hv,Aidala:2012mv,Anselmino:1994gn,Bass:2004xa}. 
The fraction 
of the spin of the proton carried by its quarks currently stands 
at~\cite{Alexakhin:2006vx} $33 \pm 3 \pm 5$ \%, 
if one relies on SU(3) symmetry for the octet 
axial charge, $g_A^8$. This is a dramatic suppression with respect 
to the value of 100 \% expected in a naive quark model or even the 65 \% 
expected in a relativistic quark model. It increases only marginally, 
to $36 \pm 3 \pm 5$ \%, if $g_A^8$ is reduced by 20 \%, as suggested by 
model calculations~\cite{Bass:2009ed} and a recent lattice 
simulation~\cite{QCDSF:2011aa}.

A number of possible theoretical explanations have been offered, 
ranging from a key role for the axial 
anomaly~\cite{Altarelli:1988nr,Carlitz:1988ab,Bodwin:1989nz,Bass:1991yx,Efremov:1989sn,Jaffe:1989jz,Shore:1991dv}
to the effect of gluon exchange 
currents~\cite{Myhrer:1988ap,Tsushima:1988xv,Hogaasen:1988},  
the effects of chiral symmetry~\cite{Schreiber:1988uw,Wakamatsu:1989qd} and, 
in the light 
of insights gained from lattice QCD studies  
a combination of both 
of these effects~\cite{Myhrer:2007cf}.
The relatively small values of $\Delta G$, the gluon spin in the proton, 
{}found in both fixed target and collider 
experiments~\cite{Adare:2007dg,Abelev:2007vt}, 
have eliminated the possibility that the axial anomaly alone might 
explain the observed suppression, although its effect may still be 
quantitatively significant.

It is clearly of great interest to find new ways to shed light on the 
origin of this remarkable phenomenon. Considerable attention is being directed 
at the measurement and interpretation of the generalised parton 
distributions (GPDs)~\cite{Belitsky:2005qn}. The moments of these GPDs can be 
related to the quark angular momentum~\cite{Ji:1996nm}, although there is 
a lively debate over the physical interpretation of those 
moments~\cite{Chen:2008ag,Ji:2010zza}. Studies of transverse momentum 
distributions (TMDs) may also offer insight into the orbital angular 
momentum carried by quarks in the nucleon~\cite{Lorce:2011kn}. In parallel 
with these experimental plans, lattice QCD has reached a level of 
sophistication such that one can relatively accurately determine the 
low moments of the GPDs which yield the total angular momentum carried by 
various flavors of quarks in the 
proton~\cite{Bratt:2010jn,Alexandrou:2011nr}, $J^{u,d,s}$.
The comparison of these results with the predictions of quark models  
after QCD evolution~\cite{Thomas:2008ga,Altenbuchinger:2010sz} 
appears promising.

In this Letter we exlore a fascinating new line of investigation into 
the spin puzzle. In particular, we extract the quark spin content 
of the octet baryons from recent lattice QCD calculations.
These are compared with the predictions of   
a relativistic quark model~\cite{Thomas:1982kv,Theberge:1980ye,Thomas:1981vc}
which includes gluon exchange 
currents~\cite{Myhrer:1988ap,Hogaasen:1995nn,Tsushima:1988qj} 
and the meson cloud required by chiral 
symmetry~\cite{Schreiber:1988uw,Kubodera:1989rc}. 
The variation of the suppression of the fraction of the spin carried 
by quarks across the octet is striking and within the relatively large 
uncertainties in the lattice results (which should improve significantly 
in the near future) this is reproduced by the model.

The lattice QCD calculations of the moments of the 
spin-dependent PDFs used here were 
based upon simulations involving $2 \, + \, 1$ flavors of 
dynamical quarks, using the Symanzik 
improved gluon action and non-perturbatively ${\cal O}(a)$ improved 
Wilson fermions~\cite{Cundy:2009yy,Bietenholz:2010jr}. 
These simulations were performed on a $24^3 \times 48$ volume with 
lattice spacing $a= 0.083 \pm 3$ fm; more details of the specific simulation 
may be found in Refs.~\cite{Bietenholz:2010jr,Horsley:2010th}. 

Because these simulations were performed at $\pi$ and $K$ masses between 
(334,460) and (401,463) MeV respectively, it is necessary to extrapolate 
them to the physical masses. This problem of extrapolation 
has been studied in great detail 
in the literature~\cite{Arndt:2001ye,Diehl:2006js,Dorati:2007bk,Chen:2001pva}, 
with the generalization to include charge symmetry breaking reported 
most recently in Ref.~\cite{Shanahan:2013xw}. Our analysis 
follows that of Ref.~\cite{Shanahan:2013xw}, where the 
spin dependent moment relevant to the proton spin problem is:
\be
\Delta q_B \equiv \langle 1 \rangle^B_{\Delta q} =  \int^1_0 dx  
(\Delta q^B(x) +  \Delta \overline{q}^B(x)) \, ,
\label{Eq:zeroth_moment}
\ee
corresponding to the matrix element in the proton of the twist-2 
operator:
\be
\mathcal{O}_{\Delta q}^{\mu}  =  \overline{q} 
\gamma_5 \gamma^{\mu} 
q \, .
\label{Eq:OPE}
\ee

To lowest order in an SU(3) expansion the matrix element of the 
operator in Eq.~(\ref{Eq:OPE}) in a member 
of the baryon octet can be expressed in terms 
of three coefficients, $\Delta \alpha, \Delta \beta$ and $\Delta \sigma$:
\begin{eqnarray}
&& \langle B (\vec{p}) | \,  \mathcal{O}_{\Delta q}^{\mu} 
 \, | B (\vec{p}) \rangle = \\ \nonumber
&& \left[
\Delta\alpha ( \overline{B} S^{\mu} B \lambda_q ) + \Delta\beta 
( \overline{B} S^{\mu} \lambda_q B) \right. \\ \nonumber
&& \left. + \Delta\sigma (\overline{B} S^{\mu} B) \textrm{Tr} 
(\lambda_q)
\right] \, .
\label{Eq:fitformula}
\end{eqnarray}
Here we have defined
\begin{equation}
\lambda^q = \frac{1}{2} \left(\xi \overline{\lambda}^q \xi^\dagger 
+ \xi^\dagger\overline{\lambda}^q \xi \right) \, ,
\end{equation}
with $\xi$ the usual $3 \times 3$ matrix constructed from the 
$\pi, \eta$ and $K$ pseudo-Goldstone bosons: 
\begin{equation}
\Phi = \frac{1}{\sqrt{2}}\left(
\begin{array}{ccc}
\frac{\pi^0}{\sqrt{2}}+\frac{\eta}{\sqrt{6}} & \pi^+ & K^+ \\
\pi^- & \frac{\eta}{\sqrt{6}} - \frac{\pi^0}{\sqrt{2}} & K^0 \\
K^- & \overline{K}^0 & -\sqrt{\frac{2}{3}} \eta \\
\end{array}
\right),\\
\end{equation}
and
\begin{equation}
\Sigma =\textrm{exp} \left(\frac{2 i \Phi}{f} \right) = \xi^2 \, . \\
\end{equation}
In addition, the octet baryon tensor, $B_{abc}$, is defined through
\begin{equation}
\textbf{B} = \left(
\begin{array}{ccc}
\frac{\Lambda}{\sqrt{2}} + \frac{\Sigma^0}{\sqrt{6}} & \Sigma^+ & p \\
\Sigma^- & \frac{\Lambda}{\sqrt{6}}-\frac{\Sigma^0}{\sqrt{2}} & n \\
\Xi^- & \Xi^0 & -\sqrt{\frac{2}{3}} \Lambda \\
\end{array}
\right),
\end{equation}
with
\begin{equation}
B_{abc}=\frac{1}{\sqrt{6}}\left(\epsilon_{abd}\textbf{B}_{dc} 
+ \epsilon_{acd} \textbf{B}_{db} \right) \, .
\end{equation}
Finally, 
$S^\mu$ is the dimensionless spin operator, with 
$\overline{B}\gamma_5\gamma^\mu B=-2\overline{B} S^\mu B$.

\begin{figure}[t]
\centering\includegraphics[width=\columnwidth,clip=true,angle=0]{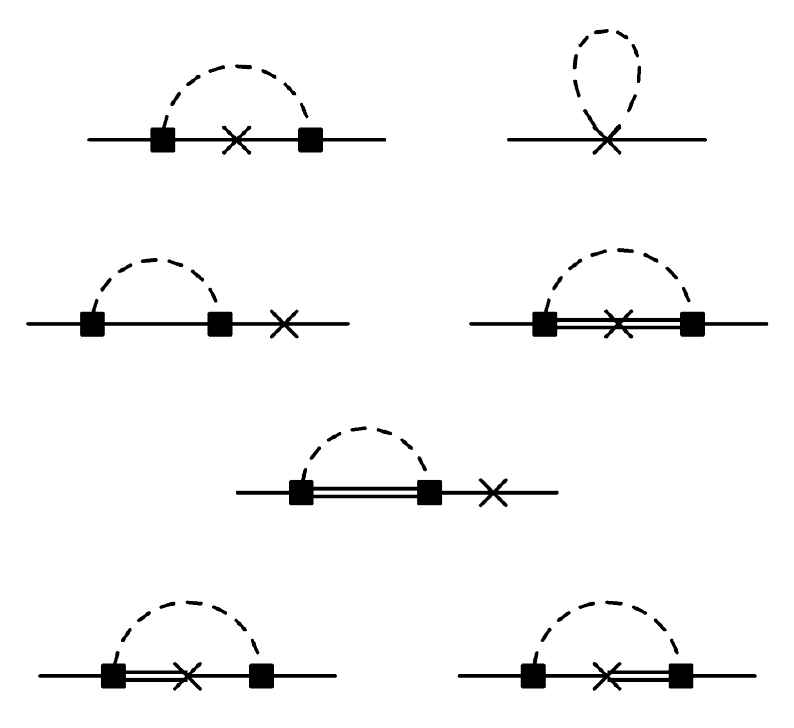}
\caption{Chiral loops included in the present calculation. The dashed 
lines represent the octet of pseudo-Goldstone bosons, the lines (double lines) 
octet (decuplet) baryons and the cross the external spin current.
\label{fig:1}}
\end{figure}

\begin{figure}[t]
\centering\includegraphics[width=\columnwidth,clip=true,angle=0]{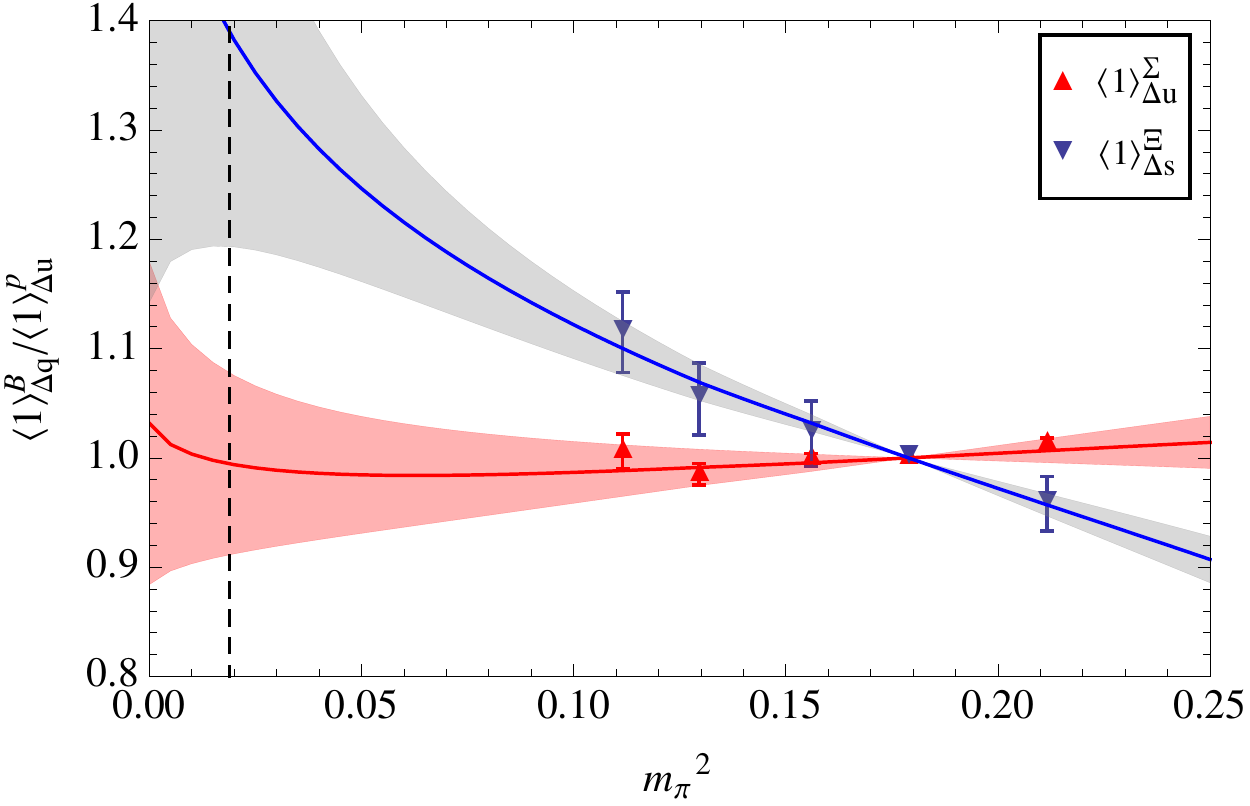}
\caption{Illustration of the fit to the ratio of the 
lattice moment of the doubly 
represented valence quarks ($u$ in the proton and $\Sigma^+$, $s$ in 
the $\Xi$) and in the $\Sigma$ and $\Xi$ hyperons to that 
in the nucleon. The vertical dashed line indicates the physical pion mass.}
\label{fig:2}
\end{figure}

The chiral corrections involving $\pi, \eta$ and $K$ loops, which 
are non-analytic in the quark mass, are illustrated 
in Fig.~\ref{fig:1}.
%
These appear at the next order in the formal expansion in quark masses. 
We include both octet and decuplet intermediate states (the latter indicated 
by double lines), while allowing for the mass difference between 
them in the numerical work. As explained in detail in 
Ref.~\cite{Shanahan:2013xw}, the inclusion of these loops adds six new 
$\mathcal{O} (m_q)$ fitting parameters, $\Delta n^{(0)}_{i=(1,6)}$, 
in addition to 
$\Delta \alpha, \Delta \beta$ and $\Delta \sigma $. These are fit 
to the 24 available lattice 
data points~\cite{Horsley:2010th,Cloet:2012db,private,Bietenholz:2011qq}, 
with the quality of the fit illustrated in Fig.~\ref{fig:2}.
%
%

Unfortunately, as there are no lattice calculations for the $\Lambda$ hyperon,  
we are unable to present results in that case. However, for the 
other members of the octet one can sum the values for 
$\Delta u_H, \Delta d_H$ 
and $\Delta s_H$ to obtain the spin fractions carried by the quarks 
in each octet baryon. Note that because the analysis of the renormalization 
of the lattice operators is not yet complete, the absolute values 
of the spin fractions are not known. However, one {\em can} 
compute the ratios of the spin fractions for the $\Sigma$ and $\Xi$ 
to that of the nucleon and these values are shown in the final column 
of Table~\ref{tableSpin}.
\begin{table*}[t]
\begin{center}
\begin{tabular}{l|c|c|c||c||c||c||}
\hline
\hline
          &MIT Bag &MIT Bag + OGE &MIT Bag + M. Cloud &MIT Bag + OGE + M. Cloud & Model & Lattice\\
\hline
$N$       &65.4 &53.8 &51.9 &43.8 & 1.0 & 1.0 \\
$\Lambda$ &77.1 &67.3 &66.4 &58.9 & 1.35 (1.33)  & - \\
$\Sigma$  &61.5 &50.8 &50.5 &42.6 & 0.97 (0.98) & 0.92 (13) \\
$\Xi$     &80.9 &72.3 &72.0 &65.2 & 1.49 (1.44) & 1.61 (33) \\
\hline
\hline
\end{tabular}
\end{center}
\caption{Spin fraction (in \%) carried by the valence quarks as the 
corrections discussed in the text are added. The column ``Model'' 
summarises the full prediction of the ratio of the spin fraction 
for each hyperon to that of the nucleon (the value in brackets corresponds 
to $R = 0.8$ fm, rather than the default 1 fm).
The final column shows the values obtained from our chiral extrapolation 
of recent lattice QCD data for the ratio of the 
quark spin in the $\Sigma$ and $\Xi$ hyperons to that in the nucleon.}
\label{tableSpin}
\end{table*}

In spite of the fact that at this stage the uncertainties are 
substantial, there is a remarkable degree of variation with 
the structure of the baryon, with the ratio of spin fractions equal 
to $0.92 \pm 0.13$ for $\Sigma:N$, while it is equal to $1.61 \pm 0.33$ 
for $\Xi:N$. These results clearly do not support the hypothesis that 
the spin suppression observed for the proton might be a 
universal property. It is therefore of 
considerable interest to investigate the predictions of models in which 
the suppression of the spin carried by quarks is dependent on structure.
With this in mind, we now apply the cloudy bag 
model (CBM), as developed in 
Refs.~\cite{Myhrer:1988ap,Schreiber:1988uw,Tsushima:1988xv,Myhrer:2007cf}, 
to this problem.

There are three major ingredients of that calculation: i) the relativistic 
suppression of the spin of a confined quark because of the orbital 
angular momentum in the lower component of its Dirac wavefunction; 
ii) color exchange current corrections associated 
with the hyperfine interaction 
mediated by one-gluon-exchange, and iii) corrections arising from the 
interchange of spin and orbital angular momentum when the cloud of 
pseudo-Goldstone bosons required by chiral symmetry is included. We 
briefly outline the calculation of each of these corrections within 
the CBM~\cite{Thomas:1982kv,Theberge:1980ye,Thomas:1981vc}.

{\em i) Relativity: \, \,}
Within the MIT bag model we take $m_s=250$ MeV as 
a representative value required 
to yield the observed hyperon masses~\cite{Myhrer:1980jy}. 
Using this value we find the spin 
suppression arising from the lower Dirac component of the $1s$ 
wave function with a bag radius $R = 1$ fm (0.8 fm) to be 0.77 (0.78), 
compared with the well-known 0.65 for a 
massless quark. Thus the most naive expectation is that there should be 
less spin suppression for the hyperons than for the nucleon.

\begin{figure}[t]
\centering\includegraphics[width=\columnwidth,clip=true,angle=0]{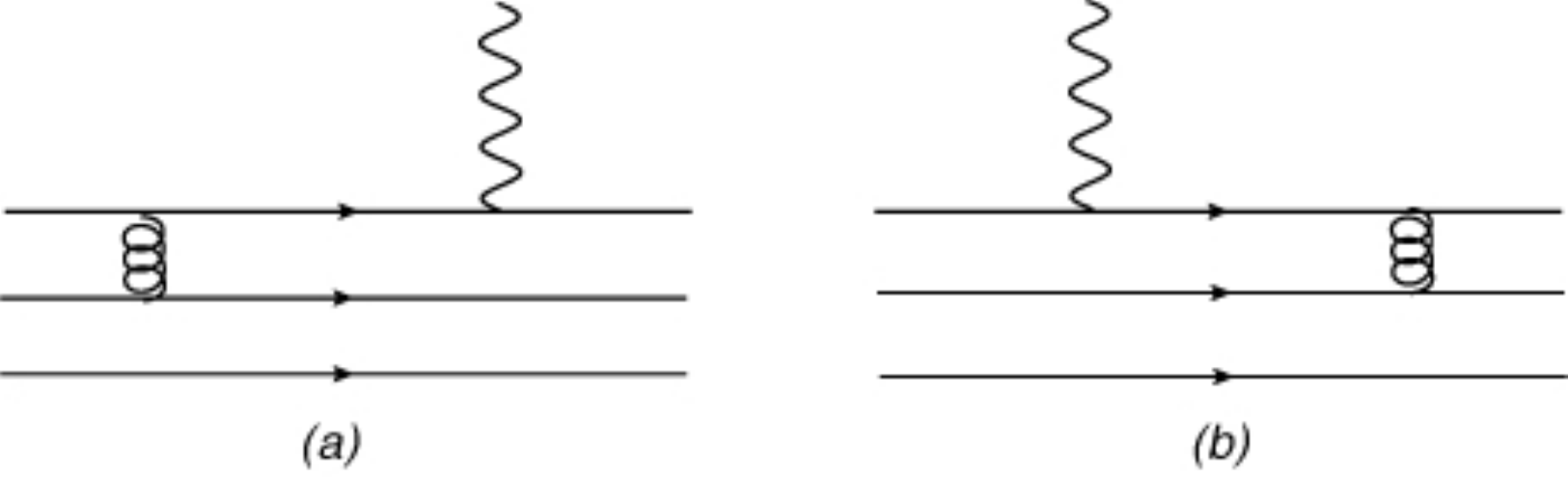}
\caption{Gluon exchange current corrections to the spin carried by 
quarks in the octet baryons. The dominant terms are those with anti-quark 
intermediate states.} 
\label{fig:3}
\end{figure}

{\em ii) Gluon exchange current correction: \, \, }
The one-gluon-exchange (OGE) force is an essential component of 
spectroscopic studies in most quark models, including the MIT bag 
model. For example, it provides a very natural explanation for 
the $\Delta$-$N$ and $\Sigma$-$\Lambda$ mass 
differences. As originally observed by 
Hogassen and Myhrer~\cite{Hogaasen:1987nj}, it also leads 
to important corrections to spin-dependent observables, such as 
magnetic moments and axial charges, through 
the processes illustrated in Fig.~\ref{fig:3}. 
%
%
The corresponding correction to the nucleon spin was calculated by 
Myhrer and Thomas~\cite{Myhrer:1988ap}, who showed that it reduced 
the quark spin content from 0.65 to around 0.5. We have repeated that 
calculation for the baryon octet. Recall that the dominant terms for a 
spin-dependent external probe are those involving an intermediate 
$q - \bar{q}$ pair (i.e., with the intermediate quark line in 
Fig.~\ref{fig:3} travelling backwards in time). In the case of the 
hyperons there are four different sets 
of matrix elements, labeled $f_{ij}$ 
with $(ij) \in (q,s)$ (with $q$ and $s$ a light or strange quark, 
respectively). The second subscript refers to the quark hit by the 
external operator and the first to the quark emitting the exchanged 
gluon. In terms of these matrix elements the corrections to the 
spins of the octet baryons, $\delta \Sigma^H$, are:
\begin{eqnarray}
\delta \Sigma^p &=& 2 f_{qq} \nonumber \\
\delta \Sigma^\Lambda &=& 2 f_{sq} \nonumber \\
\delta \Sigma^\Sigma &=& \frac{4}{3} f_{qq} - \frac{2}{3} f_{sq} 
+\frac{4}{3} f_{qs} \nonumber \\
\delta \Sigma^\Xi &=& \frac{4}{3} f_{ss} - \frac{2}{3} f_{qs} 
+\frac{4}{3} f_{sq} \, ,
\label{eqn:OGE}
\end{eqnarray}
where for $R=1$ fm (0.8 fm) our numerical evaluation gives 
$f_{qq}=-0.058 (-0.058), \, f_{sq}=-0.049 (-0.040), \, 
f_{qs}=-0.047 (-0.049), \, f_{ss}=-0.039 (-0.042)$. 
This yields the values for the spin 
{}fractions shown in the column of Table~\ref{tableSpin} labelled 
``MIT Bag + OGE''. There is only a small variation of the size of the 
correction across the octet, with values varying from 12\% in the 
proton to 8\% in the $\Xi$.

{\em iii) Chiral corrections: \, \, }
For the nucleon the correction arising from the pion cloud was first 
discussed by Schreiber and Thomas~\cite{Schreiber:1988uw}, who found  
a reduction of the spin fraction carried by quarks by 20-30\%. 
At the time there was a serious concern about potential double counting 
if one were to combine the OGE and pion corrections, which was only 
resolved (in favor of no significant double counting) 
after recent studies of the 
$\Delta$-$N$ mass splitting in quenched 
and full lattice QCD~\cite{Myhrer:2007cf,Young:2002cj}.
Already in the late 80's, Kubodera and collaborators~\cite{Kubodera:1989rc}
combined the OGE corrections with the chiral loops for pions, etas and 
kaons under the assumption that there was no double counting problem. 

We have repeated that calculation for the full 
octet of baryons, working 
strictly within the CBM and using a typical bag radius of 1 fm everywhere 
(including the CBM form factors at the meson-baryon vertices). The 
effect of the meson cloud on the MIT bag is shown in column 
``MIT Bag + M. Cloud'' of Table~\ref{tableSpin}, while the 
final results, including relativity, OGE and the meson cloud corrections, 
are shown in the column labelled
``MIT Bag + OGE + M. Cloud''. We see that once all effects have been 
included there is a substantial variation 
in the spin fractions carried by the quarks across the octet. The 
meson cloud correction is considerably smaller in the $\Xi$ than in the 
nucleon. That, combined with the less relativistic motion of the 
heavier strange quark, results in the spin fraction in the $\Xi$ being 
quite a bit larger than in the nucleon.

The next-to-last column in Table~\ref{tableSpin} shows the ratios of 
the quark spin fractions in the hyperons compared with that in the 
nucleon. That the dependence on the bag radius is minimal
is illustrated in the second last column
of Table~\ref{tableSpin}, where the number in brackets shows the model result
with the bag radius changed from 1 fm to 0.8 fm 
(the latter unrealistic for hyperons) everywhere in the calculation.
At the current level of precision there is
clearly very good agreement
between the values calculated within the CBM and those
extracted from lattice QCD.
It will be extremely interesting to 
investigate the hyperon spin fractions in other models. In addition, 
this work illustrates the importance of further work to reduce the 
statistical errors of lattice QCD simulations of the moments of hyperon 
spin dependent PDFs and to extend them to quark masses closer to the physical 
region. 

\section*{Acknowledgements}
This work was supported by the National Science Foundation (US) through 
grant no. PHY-1068305 (FM) and by  the Australian Research Council through 
funding to the ARC Centre of Excellence in Particle Physics at the 
Terascale, an 
Australian Laureate Fellowship (FL0992247, AWT), DP110101265 (RDY) and 
FT120100821 (RDY). 

\end{document}